\documentclass[twoside,11pt]{article}
\usepackage{times}
\usepackage{amsmath,amsfonts}
\usepackage{graphicx}
\usepackage{xcolor}
\usepackage{gensymb}
\usepackage{dirtytalk}
\usepackage{fullpage}
\usepackage{textcomp}
\usepackage{siunitx}
\usepackage{ccaption}
\usepackage[colorlinks = true,
          linkcolor = blue,
          urlcolor  = blue,
          citecolor = blue,
          anchorcolor = blue]{hyperref}
\usepackage[numbers,compress]{natbib}

\title{Computational analysis of a 9D model for a small DRG neuron}
\date{\vspace{-5ex}}
\author{
    Parul Verma$^1$ \qquad
    Achim Kienle$^{2,3}$ \qquad
    Dietrich Flockerzi$^{2,3}$ \\
    $^*$Doraiswami Ramkrishna$^{1}$
    \\ \\
    $^1$ Davidson School of Chemical Engineering, Purdue University, USA \\
    $^2$ Max Planck Institute for Dynamics of Complex Technical Systems, Magdeburg, Germany\\
    $^3$ Otto von Guericke University, Magdeburg, Germany\\
    \texttt{$^*$ramkrish@purdue.edu}
}

\begin{document}

\maketitle

\section*{Abstract}

Small dorsal root ganglion (DRG) neurons are primary nociceptors which are responsible for sensing pain. Elucidation of their dynamics is essential for understanding and controlling pain. To this end, we present a numerical bifurcation analysis of a small DRG neuron model in this paper. The model is of Hodgkin-Huxley type and has 9 state variables. It consists of a Na\textsubscript{v}1.7 and a Na\textsubscript{v}1.8 sodium channel, a leak channel, a delayed rectifier potassium and an A-type transient potassium channel. The dynamics of this model strongly depends on the maximal conductances of the voltage-gated ion channels and the external current, which can be adjusted experimentally. We show that the neuron dynamics are most sensitive to the Na\textsubscript{v}1.8 channel maximal conductance ($\overline{g}_{1.8}$). Numerical bifurcation analysis shows that depending on $\overline{g}_{1.8}$ and the external current, different parameter regions can be identified with stable steady states, periodic firing of action potentials, mixed-mode oscillations (MMOs), and bistability between stable steady states and stable periodic firing of action potentials. We illustrate and discuss the transitions between these different regimes. We further analyze the behavior of MMOs. As the external current is decreased, we find that MMOs appear after a cyclic limit point. Within this region, bifurcation analysis shows a sequence of isolated periodic solution branches with one large action potential and a number of small amplitude peaks per period. For decreasing external current, the number of small amplitude peaks is increasing and the distance between the large amplitude action potentials is growing, finally tending to infinity and thereby leading to a stable steady state. A closer inspection reveals more complex concatenated MMOs in between these periodic MMOs branches, forming Farey sequences. Lastly, we also find small solution windows with aperiodic oscillations, which seem to be chaotic. The dynamical patterns found here as a function of different parameters contain information of translational importance as their relation to pain sensation and its intensity is a potential source of insight into controlling pain.

\section{Introduction}
\label{sec:intro}
Neurons display a variety of rich dynamics such as repetitive firing of action potentials, bursting, mixed-mode oscillations, and bistability. The diversity of dynamics displayed by a multitude of neurons has led several researchers to bifurcation theory to understand the transition from one dynamical pattern to the other for more than 40 years~\citep{troy1978bifurcation,hassard1978bifurcation,rinzel/1980a,izhikevich2007dynamical}. Starting with the analysis of low dimensional models such as Hodgkin-Huxley and Fitzhugh-Nagumo equations, it has recently been used for higher dimensional models such as a 14D model of a pyramidal cell~\citep{ju2018torus}, as well. In this paper, we employ numerical bifurcation analysis to understand the dynamics of a 9D model of a small dorsal root ganglion (DRG) neuron.  

Small DRG neurons are primary nociceptors, i.e., they are responsible for sensing pain \citep{sherrington1903qualitative}. From a theoretical point of view, pain corresponds to repetitive firing of action potentials~\citep{djouhri2006spontaneous,dubin2010nociceptors}. To understand how pain can be controlled, it is therefore essential to determine how the transition to periodic firing of action potentials depends on the physiological parameters and how these parameters can be manipulated in a suitable way.

While limited numerical~\citep{sheets2007nav1,choi2011physiological,sundt2015spike} and extensive experimental~\citep{amir1999drg,rush2007drg,krames2014drg,berta2017drg} studies have been executed for this type of neuron, a detailed bifurcation analysis has not been undertaken so far. The importance of using bifurcation theory to understand pain is emphasized in the works of~\cite{rho2012identification} and~\cite{ratte2014criticality} where 2D and 3D models of an afferent sensory neuron were analyzed with regard to neuropathic pain and, subsequently, bifurcation theory aided in finding parametric regions of pain and no-pain. Previous work on a model of a small DRG neuron~\citep{Verma757187} also illustrates the utility of bifurcation theory for understanding pain. In that paper, genetic mutations in sodium channels that are associated with pain sensation were also investigated.

In the present paper, we use the aforementioned theory extensively to find the bifurcations explaining the excitability patterns of this model. We perform both one-parameter and two-parameter continuation of model solutions, with external current as the primary bifurcation parameter and maximal conductance of one of the voltage-gated ion channels as the secondary parameter. Here, we find different solution regimes consisting of stable steady states, periodic firing of action potentials, and mixed-mode oscillations (MMOs). The latter are periodic or aperiodic oscillatory solutions consisting of small amplitude (subthreshold) and large amplitude (action potential) peaks. They have been recorded in DRG neuron cultures before~\citep{amir1999drg,zheng2012enhanced}. Besides, they have been observed in many other chemical and neuronal systems~\citep{brons2008mmoreview,desroches2012mixed}, and are therefore of broader interest. We elaborate on the mechanisms of onset and disappearance of MMOs, and compare them to other extensively analyzed MMO-generating systems.

This paper is organized as follows. In Sec.~\ref{sec:Model}, we describe the model and show various patterns of behaviour using dynamic simulation for selected parameter values. Sec.~\ref{sec:bif} identifies different parameter regions corresponding to the different patterns of behavior using one- and two-parameter continuation with external current as the primary bifurcation parameter and maximal conductance of the Na\textsubscript{v}1.8 sodium channels as the secondary bifurcation parameter. To account for model uncertainties, we also study sensitivity with respect to the other model parameters afterwards using two-parameter continuation of critical boundaries. In Sec.~\ref{sec:mmo}, our focus is on MMOs. We use periodic continuation to calculate a sequence of periodic solution branches with one large amplitude action potential and various numbers of small amplitude subthreshold peaks per period. Further, dynamic simulations illustrate the existence of more complex concatenated periodic and aperiodic MMOs. Lastly, in Sec.~\ref{sec:discussions}, we discuss and conclude our results, providing remarks on problems that still need to be addressed.

\section{Model description and first simulation results}
\label{sec:Model}
In this paper, we focus on a single compartment minimal conductance model. Following~\cite{choi2011physiological}, the model accounts for currents due to two sodium channels: Na\textsubscript{v}1.7 ($I_{1.7}$) and Na\textsubscript{v}1.8 ($I_{1.8}$), two potassium channels: a delayed rectifier ($I_{K}$) and an A-type transient ($I_{KA}$) channel, and a leak channel ($I_l$). These are the primary ion channels found on the membrane of a small DRG neuron. The equation for membrane voltage dynamics are written in the following Hodgkin-Huxley~\cite{HodgkinHuxley1952} type of form:
\begin{equation}
C\frac{\mathrm{d}V}{\mathrm{d}t} = \frac{I_{ext}}{A} - (I_{1.7} + I_{1.8} + I_{K} + I_{KA} + I_{l}),
\label{eqak1}
\end{equation}
where $V$ is the neuron membrane voltage (\si{mV}), $C$ is the specific membrane capacitance (\si{\mu F/cm^2}), $t$ is time (\si{ms}, milliseconds), $I_{ext}$ is the external applied current, and $A$ is the area. $I_{ext}/A$ in Eq.~\eqref{eqak1} has the dimension \si{\mu A/cm^2}. The other specific currents on the right hand side of Eq.~\eqref{eqak1} are calculated as follows: $I_{1.7} = \overline{g}_{1.7} \:  m_{1.7}^3 \:  h_{1.7} \:  s_{1.7} \:  (V-E_{Na})$, $I_{1.8} = \overline{g}_{1.8} \: m_{1.8} \: h_{1.8} \: (V-E_{Na})$, $I_{K} = \overline{g}_K \: n_K \: (V-E_K)$, $I_{KA} = \overline{g}_{KA} \: n_{KA} \: h_{KA} \: (V-E_{K})$, and $I_{l} = \overline{g}_{l} \: (V-E_{l})$ . Therein, $\overline{g}_i$ and $E_j$ are the specific maximal conductances (\si{mS/cm^2}) and equilibrium potentials (\si{mV}), respectively, for $i = 1.7, 1.8, K,\\ KA, l$, and $j = Na, K, l$. 
 
All the activation and inactivation variables $x$ ($x = m_{1.7}, h_{1.7}, s_{1.7}, m_{1.8}, h_{1.8}, n_K, n_{KA}, h_{KA}$) are dimensionless variables that can vary between 0 to 1. They are calculated from a corresponding differential equation of the following form:
\begin{equation}
    \frac{\mathrm{d}x}{\mathrm{d}t} = \frac{x_{\infty} (V) - x}{\tau_x (V)},
\end{equation}
where the nonlinear expressions for $x_{\infty}$ and $\tau_x$ are given in the Appendix~\ref{sec:appendix}.

Leak current kinetics, area, membrane capacitance, and equilibrium potential values for a small DRG neuron were extracted from~\cite{choi2011physiological}. Maximal conductances $\overline{g}_K$ and $\overline{g}_{KA}$ were estimated to ensure that their corresponding currents were 6 \si{nA} and 1 \si{nA} at 0 \si{mV} when the cell is initially depolarized to -120 \si{mV}, and $\overline{g}_{1.7}$ was set to 18 \si{mS/cm^2}, based on~\cite{choi2011physiological}. $\overline{g}_{1.8}$ was set to 7 \si{mS/cm^2}, which led to the generation of one action potential (current threshold) at 100 \si{pA} when the neuron is at the resting membrane potential (RMP), which is determined by simulating the model for $I_{ext} = 0$ \si{pA}. The current threshold of 100 \si{pA} was chosen based on approximate values from previous experiments and simulations~\citep{sheets2007nav1,zheng2012enhanced}. The parameter values of this model are listed in Table~\ref{tab:1}. These parameter values result in an RMP of -66.48 \si{mV} which belongs to the physiological range of RMP recorded in small DRG neurons in~\cite{huang2018atypical}, and the resulting action potential amplitude (approximately 120 \si{mV}) is comparable to that reported in~\cite{yang2016nav17}.

\begin{table}
\caption{Model parameter values}
\label{tab:1}       
\begin{tabular}{lll}
\hline\noalign{\smallskip}
Parameter & Value & Units  \\
\noalign{\smallskip}\hline\noalign{\smallskip}
$A$ (area) & 2168.00 & \si{\mu m^2} \\
$C$ & 0.93 & \si{\mu F/cm^2} \\
$E_{Na}$ & 67.10 & \si{mV} \\
$E_{K}$ & -84.70 & \si{mV} \\
$E_{l}$ & -58.91 & \si{mV} \\
$\overline{g}_{1.7}$ & 18.00 & \si{mS/cm^2} \\
$\overline{g}_{1.8}$ & 7.00 & \si{mS/cm^2} \\
$\overline{g}_K$ & 4.78 & \si{mS/cm^2} \\
$\overline{g}_{KA}$ & 8.33 & \si{mS/cm^2} \\
$\overline{g}_{l}$ & 0.0575 & \si{mS/cm^2} \\
\noalign{\smallskip}\hline
\end{tabular}
\end{table}

The final equation (\ref{eqak1}) reads:
\begin{align*}
    C \frac{\mathrm{d}V}{\mathrm{d}t} = \frac{I_{ext}}{A} & -(\overline{g}_{1.7} m_{1.7}^3 h_{1.7} s_{1.7} (V-E_{Na})
    \\
    & + \overline{g}_{1.8} m_{1.8} h_{1.8} (V-E_{Na})
    \\
    & + \overline{g}_K n_K (V-E_K)
    \\
    & + \overline{g}_{KA} n_{KA} h_{KA} (V-E_K) 
    \\
    & + \overline{g}_l (V-E_l))
\end{align*}

Numerical integration and bifurcation analysis were primarily done in XPPAUT~\citep{ermentrout2002simulating} and cross checked with MATCONT~\citep{matcont}. In XPPAUT, default settings were used except for the following: \texttt{NTST} = 100, \texttt{Method} = Stiff, \texttt{Tolerance} = 1e-7, \texttt{EPSL}, \texttt{EPSU}, \texttt{EPSS} = 1e-7, \texttt{ITMX}, \texttt{ITNW} = 20, \texttt{PARMIN} = 0, \texttt{PARMAX} = 300. In MATCONT, the following settings were kept: \texttt{MaxCorrIters} = 20, \texttt{MaxTestIters} = 20, \texttt{FunTolerance} = 1e-6, \texttt{VarTolerance} = 1e-7, \texttt{TestTolerance} = 1e-7, \texttt{NTST} = 300, \texttt{tolerance} = 1e-4, \texttt{MaxStepsize} = 1 for steady state continuation and 10 for periodic solution continuation. Integration was performed using ode15s. Integration option \texttt{RelTol} was set to 1e-8.

\begin{figure*}
  \includegraphics[width=1\textwidth]{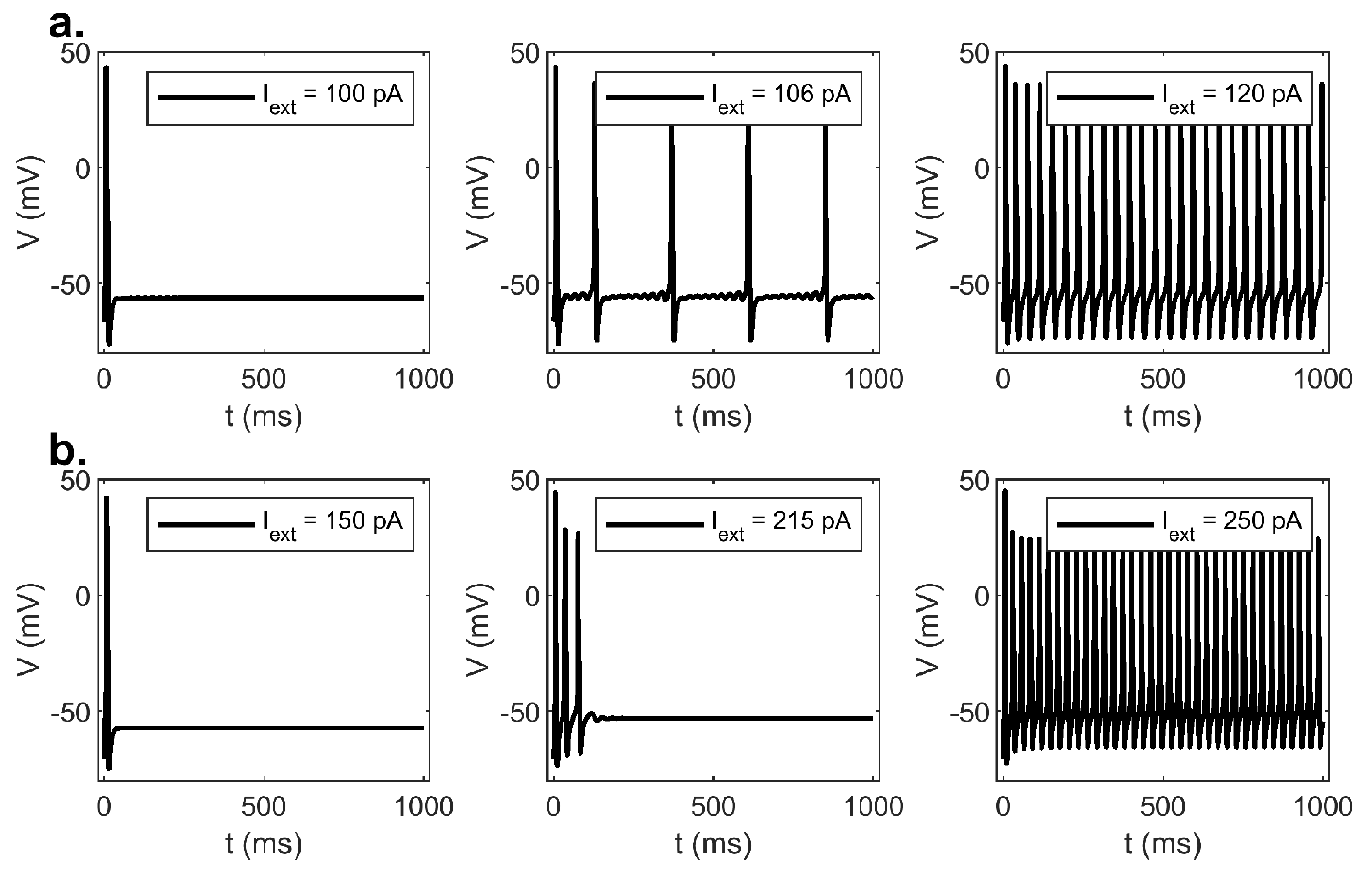}
\caption{Dynamic simulations of action potentials. For higher values of $\overline{g}_{1.8}$, MMOs are observed. a.: Dynamic simulations for $\overline{g}_{1.8}$ at 7 \si{mS/cm^2}, and $I_{ext}$ = 100, 106, 120 \si{pA}.  b.: Dynamic simulations for $\overline{g}_{1.8}$ at 4.5 \si{mS/cm^2}, and $I_{ext}$ = 115, 215, 230 \si{pA}. No MMOs are observed in this case.}
\label{fig:1}       
\end{figure*}

In a first step, we present selected dynamic simulations of the above equations to illustrate some characteristic patterns of behavior, to be analyzed in more detail in the remainder. Results are shown in Fig. \ref{fig:1}. Initial condition for all simulations was the stable steady state for $I_{ext}=0$ \si{pA}.

In the first row of Fig.~\ref{fig:1} the maximum conductance of the Na\textsubscript{v}1.8 channel equals 7 \si{mS/cm^2} and the value of the external current is increased from 100 \si{pA} in the left diagram, to 106 \si{pA} in the middle, to 120 \si{pA} in the right diagram. In the left diagram, for the lowest value of $I_{ext}$, a stable steady state is attained after the firing of an action potential, whereas periodic firing of large amplitude action potentials is observed for the highest value of $I_{ext}$ in the right diagram. For values of $I_{ext}$ in between, there is a region where mixed mode oscillations (MMOs) are observed as illustrated in the middle diagram. There, after some initial transient, a periodic regime is attained with one large amplitude action potential and eight small amplitude subthreshold peaks per period. 

A different situation is seen in the second row of Fig.~\ref{fig:1} with some representative simulations. There, the maximum conductance of the Na\textsubscript{v}1.8 channel is equal to 4.5 \si{mS/cm^2}. Again the values of the external current are increasing from the left to the right. The qualitative behavior in the left and the right diagrams, for the lowest and the highest value of $I_{ext}$, is similar to the behavior shown in the corresponding diagrams of Fig.~\ref{fig:1}a. However, in contrast to Fig.~\ref{fig:1}a, no MMOs are found in the intermediate range of $I_{ext}$. We illustrate this in the middle diagram for one specific value of $I_{ext}$ of 215 \si{pA}, where the cell potential decays to a stable steady state after firing of three action potentials. As we will show in more detail in the next section, MMOs do not exist for any value of the intermediate range for $\overline{g}_{1.8}$ equal to 4.5 \si{mS/cm^2}.

Both of the cases shown in Fig.~\ref{fig:1} have been observed in DRG culture recordings. See~\cite{huang2018atypical} for recordings resembling Fig.~\ref{fig:1}b, and~\cite{amir1999drg} and~\cite{zheng2012enhanced} for recordings displaying MMOs as in Fig.~\ref{fig:1}a.

\section{Numerical bifurcation analysis} 
\label{sec:bif}
\begin{figure*}
  \includegraphics[width=1\textwidth]{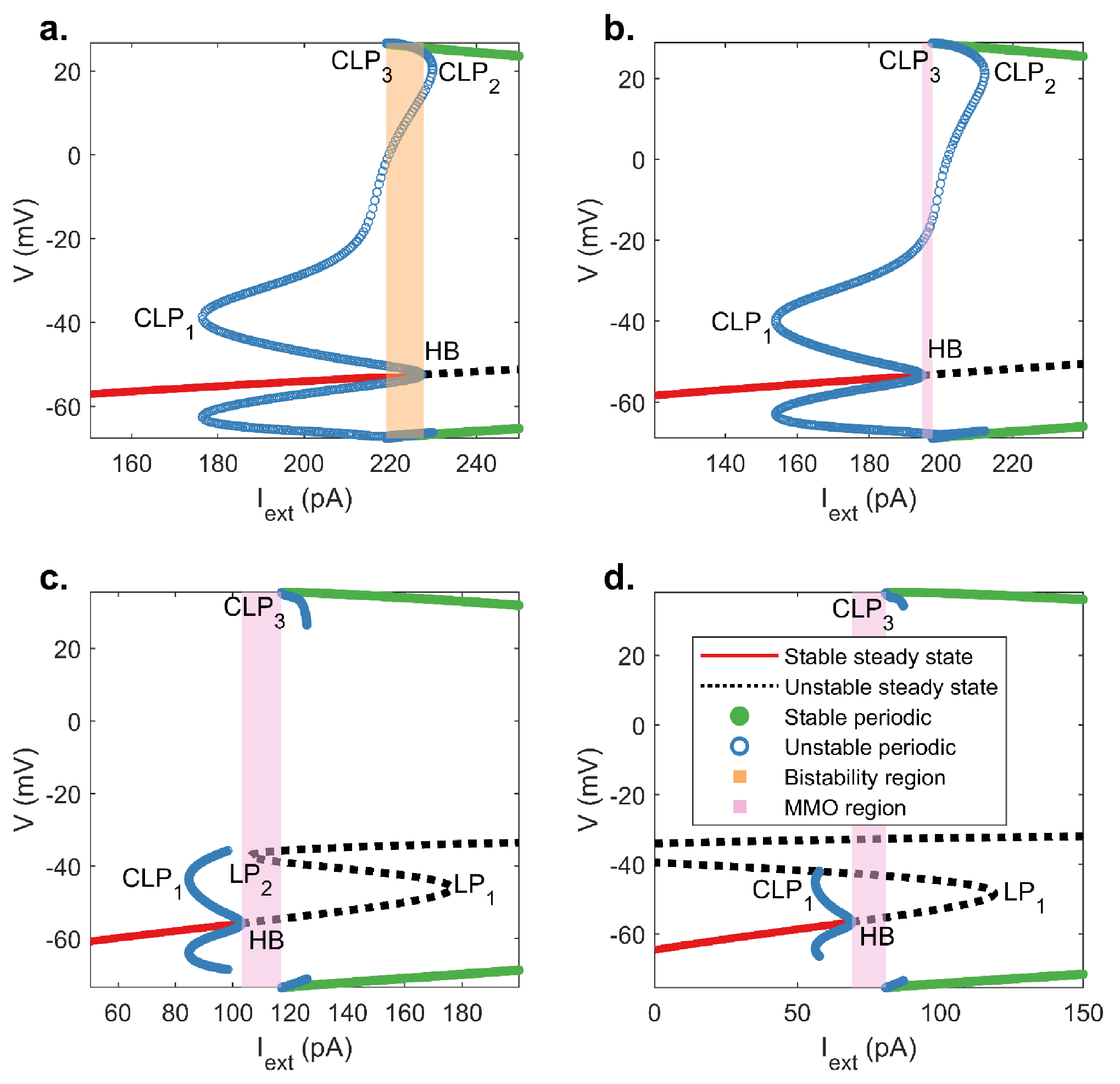}
\caption{Bifurcation diagrams for $\overline{g}_{1.8}$ = 4.5, 5, 7 and 8 \si{mS/cm^2} for diagrams a, b, c and d, respectively. For lower values of $\overline{g}_{1.8}$ in diagram (a), MMOs are not observed, and there is a region of bistability between steady state and periodic firing of action potentials, as shown by the orange shaded region. This bistability is not present in diagrams b, c, d. Instead, MMOs are observed in these diagrams in the purple shaded region. MMOs solution branches will be discussed separately in section 3 and are not included in this figure.
HB: Hopf bifurcation point, CLP: Cyclic limit point, LP: limit point}
\label{fig:2}
\end{figure*}

In order to explain the transitions between different dynamical patterns observed in this model, we perform one-parameter and two-parameter continuation of solutions, with $I_{ext}$ as the primary bifurcation parameter, and $\overline{g}_{1.8}$ as the secondary bifurcation parameter. First, we perform one-parameter continuations of steady state and periodic solutions upon varying the primary bifurcation parameter $I_{ext}$. Results are shown in Fig.~\ref{fig:2} for four different values of $\overline{g}_{1.8}$. The first diagram in Fig.~\ref{fig:2}a is for a value of 4.5 \si{mS/cm^2} corresponding to the scenario in Fig.~\ref{fig:1}b, whereas the third diagram in Fig.~\ref{fig:2}c is for a value of 7 \si{mS/cm^2} corresponding to the scenario in Fig.~\ref{fig:1}a. Two additional scenarios for values of 5 and 8 \si{mS/cm^2} are shown in Figs.~\ref{fig:2}b and \ref{fig:2}d.

In all the four diagrams of Fig.~\ref{fig:2}, a branch of stable steady states is obtained for low values of $I_{ext}$ starting from the left boundary of the corresponding diagram. It is indicated by the red solid line and correspond to the behavior shown in the left diagrams of Figs.~\ref{fig:1}a, b. The stable steady states become unstable at a subcritical Hopf bifurcation point (HB), from where a branch of unstable periodic solutions emerges indicated by the blue circles in Fig.~\ref{fig:2}. 

Furthermore, in all the four diagrams of Fig.~\ref{fig:2}, a branch of stable periodic solutions is observed for high values of $I_{ext}$ at the right boundary of the corresponding diagrams. It is indicated by the green filled circles and correspond to periodic firing of action potentials as shown in the right diagrams of Figs.~\ref{fig:1}a, b. In all the four cases, these branches of stable periodic solutions lose their stability at a cyclic limit point (CLP$_3$), giving rise to a branch of unstable periodic solutions.

Qualitative differences in the four diagrams of Fig.~\ref{fig:2} occur with respect to the unstable periodic branches indicated by the blue circles. In the first two diagrams, the unstable periodic solution branches are connected and show two more cyclic limit points (CLP$_1$ and CLP$_2$); after the third cyclic limit point (CLP$_3$), they turn into the stable periodic solution branch with periodic firing of action potentials as described above. In contrast to this, in the last two diagrams, the unstable periodic solution branches are disconnected and CLP$_2$ disappears. At the end points of these branches, the period of the unstable periodic solutions is increasing rapidly during continuation with XPPAUT and MATCONT, indicating the presence of a period infinity solution at the end points.

Another difference occurs with respect to the unstable steady state branches indicated by the dashed lines in Fig.~\ref{fig:2}. They display hysteresis with two limit points (LP$_1$ and LP$_2$) in Fig.~\ref{fig:2}c, one of which (LP$_2$) has moved out of the positive range for $I_{ext}$ in Fig.~\ref{fig:2}d.

Further, we see that the $I_{ext}$ value of the bifurcation points varies significantly between the four diagrams of Fig.~\ref{fig:2}, indicating a high sensitivity to $\overline{g}_{1.8}$. For the increasing values of $\overline{g}_{1.8}$ from Fig.~\ref{fig:2}a to~\ref{fig:2}d, the bifurcation points are shifted to lower values of $I_{ext}$. Besides the absolute $I_{ext}$ value, we find that the relative position of the HB point and the CLP$_3$ point is of major importance for the qualitative differences reported in Fig.~\ref{fig:1}. In Fig.~\ref{fig:2}a, the $I_{ext}$ value of HB is higher than that of CLP$_3$, leading to an overlap between stable steady state and stable periodic solutions indicated by the orange region of Fig.~\ref{fig:2}a. An increase of $I_{ext}$ will lead to the periodic firing of action potentials when the HB point is crossed as shown in the scenario in Fig.~\ref{fig:1}b. A transition back to stable steady states will occur at the value of the CLP$_3$ point if $I_{ext}$ is decreased again afterwards. Between the CLP$_3$ and the HB point in Fig.~\ref{fig:2}a, the system is bistable, i.e., the initial conditions regulate whether a stable periodic or a stable steady state solution is attained. 

The situation is fundamentally different in Figs.~\ref{fig:2}c and~\ref{fig:2}d. Here, the $I_{ext}$ value of the HB point is lower than the value of the CLP$_3$ point, leading to a situation where no stable attracting solutions are shown in the purple shaded region of these diagrams. This is the region where various types of stable periodic and aperiodic MMOs exist. The MMOs solution branches are missing in Fig.~\ref{fig:2} and will be discussed in the next section. Further, we will show that in these cases, the CLP$_3$ point provides a strict upper limit of the MMOs region, whereas the lower limit is not determined by the HB point but by a value close by where the time between subsequent large amplitude action potentials of the MMOs tends to infinity.

\begin{figure*}
    \centering
    \includegraphics[width=\textwidth]{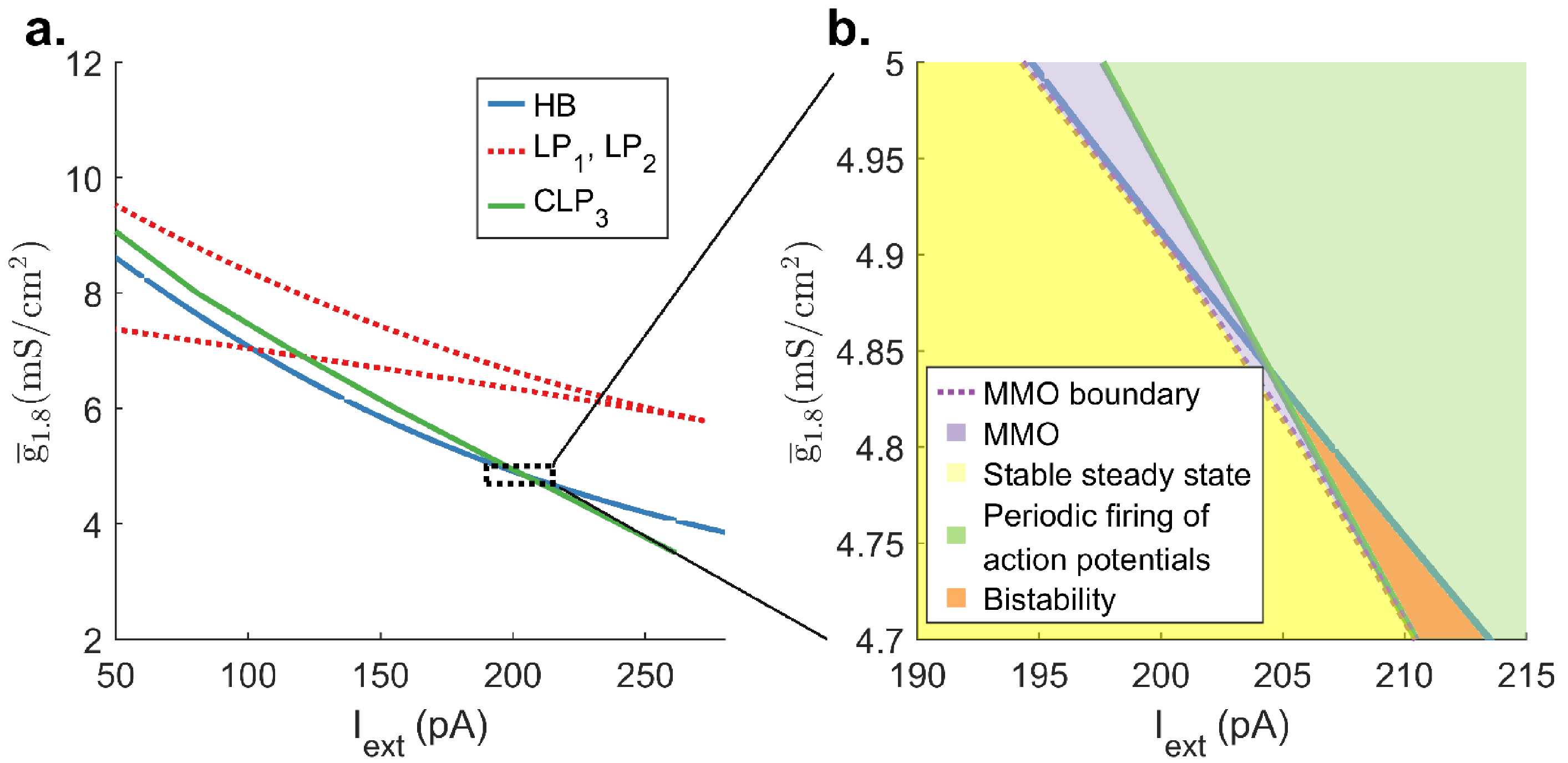}
    \caption{Two parameter plot with $\overline{g}_{1.8}$ as the secondary continuation parameter. a.: Variation over a large interval of $\overline{g}_{1.8}$. b.: Zoomed in version of a. near the intersection of the HB point and the CLP$_3$ point.}
    \label{fig:18cont}
\end{figure*}

To map out the regions in the $I_{ext}$ and $\overline{g}_{1.8}$ parameter space with different patterns of behavior, we perform a two-parameter continuation of the relevant critical points HB, CLP$_3$, LP$_1$, and LP$_2$. The results are shown in Fig.~\ref{fig:18cont}. As mentioned above, the upper boundary of the MMOs region is the curve of the CLP$_3$ points, whereas the lower boundary is a solution where the time between subsequent action potentials tends to infinity close to the HB curve. These boundaries are seen best in Fig.~\ref{fig:18cont}b. A direct calculation and continuation of the lower boundary is substantially challenging and was not done. Instead, we determine the lower boundary of the MMOs region by point-wise dynamic simulation over a prolonged time period. In summary, we find that the transition from the stable steady state region (`no pain') to the repetitive firing of action potentials (`pain') differs depending on the value of $\overline{g}_{1.8}$. For high values of $\overline{g}_{1.8}$, MMOs occur. As we will show in the next section, the frequency of action potentials in this region is increasing step by step as the stimulus $I_{ext}$ is increased. In contrast to this, for values of $\overline{g}_{1.8}$ below the intersection of the HB and the CLP$_3$ curve, we have the orange bistable region with a `hard' onset of the periodic firing of action potentials with high frequency.

This analysis suggests that the small DRG neuron dynamics depend strongly on the expression of Na\textsubscript{v}1.8. For lower expression of Na\textsubscript{v}1.8, it may not display subthreshold oscillations. This can explain the variability in DRG culture recordings reported in~\cite{amir1999drg,zheng2012enhanced,huang2018atypical}. 

\begin{figure*}
    \centering
    \includegraphics[width=\textwidth]{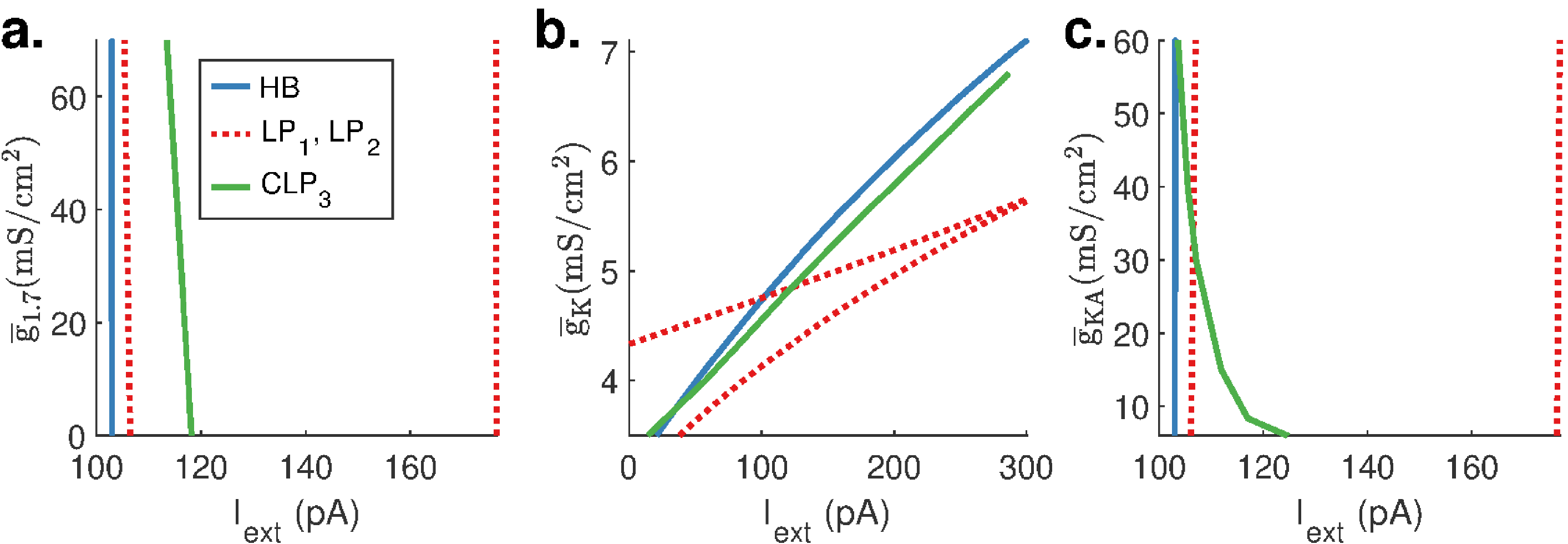}
    \caption{Two parameter plot with the following secondary continuation parameters: a.: $\overline{g}_{1.7}$, b.: $\overline{g}_K$ and c.: $\overline{g}_{KA}$.}
    \label{fig:2cont}
\end{figure*}

We also investigate the influence of the other maximal conductances $\overline{g}_i \: (i=1.7,K,KA)$ using two-parameter continuations of the critical bifurcation points with $I_{ext}$ as the primary bifurcation parameter and the corresponding maximal conductance as the secondary bifurcation parameter. Results are shown in Fig.~\ref{fig:2cont}. As seen in Fig.~\ref{fig:2cont}a and Fig.~\ref{fig:2cont}c, $\overline{g}_{1.7}$ and $\overline{g}_{KA}$ have negligible effect on the HB and the LP points. The CLP$_3$ point is sensitive only to lower values of $\overline{g}_{KA}$. All the bifurcation points are sensitive to $\overline{g}_{K}$, as seen in Fig.~\ref{fig:2cont}b. Here, the CLP$_3$, HB, and LP points vary substantially from 0 to 300 \si{pA} in a small range of $\overline{g}_{K}$.

\section{Mixed-mode oscillations}
\label{sec:mmo}
In this section, our focus is on the periodic and aperiodic MMOs solutions already mentioned in the previous sections. For the characterization of periodic MMOs solutions, we apply the nomenclature introduced, for example, in~\cite{FareyBZ}. Basic MMOs patterns consist of $L$ large amplitude peaks (action potentials) followed by $S$ small amplitude (subthreshold) peaks per period, termed as $L^S$ patterns in this notation. In particular, $L$ is equal to one in the remainder of this section. More complex patterns arise due to the concatenation of different basic patterns, for example, a pattern of the form $L_1^{S_1}L_2^{S_2}$ can occur between the basic patterns $L_1^{S_1}$ and $L_2^{S_2}$. It consists of $L_1$ action potentials, followed by $S_1$ subthreshold peaks, followed by $L_2$ action potentials, followed by $S_2$ subthreshold peaks in each period.

\begin{figure*}
    \centering
    \includegraphics[width=\textwidth]{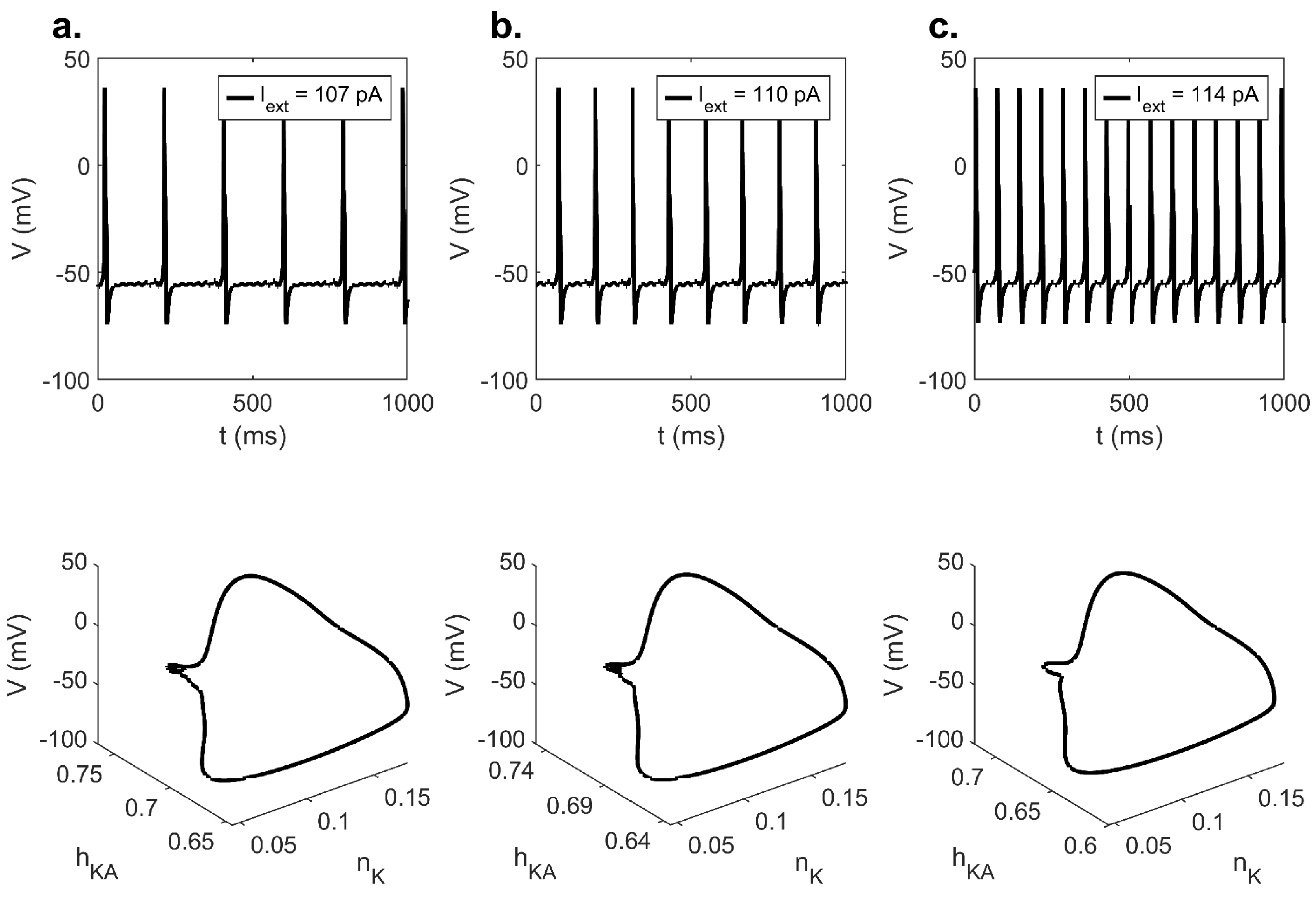}
    \caption{Basic MMOs solutions of the type: a.: $1^6$, b.: $1^3$, and c.: $1^1$ for selected values of $I_{ext}$. Upper row: temporal evolution of membrane voltage, lower row: orbits in the $V$, $h_{ka}$, $n_{K}$ phase space.}
    \label{fig:mmo1}
\end{figure*}

We show some characteristic basic patterns of MMOs in Fig.~\ref{fig:mmo1} for different values of $I_{ext}$. In the remainder of this section, we use the default parameter values from Table~\ref{tab:1} and the value of $\overline{g}_{1.8}$ is equal to 7 \si{mS/cm^2} corresponding to Fig.~\ref{fig:2}c. According to this figure and our previous results, we expect MMOs in the range of $I_{ext}$ of roughly 103 to 117 \si{pA}. More precise values will be given in the course of this discussion. According to the nomenclature mentioned above, the patterns in Fig.~\ref{fig:mmo1} can be characterized as $1^6$ for $I_{ext}$ = 107 \si{pA}, $1^3$ for $I_{ext}$ = 110 \si{pA}, and $1^1$ for $I_{ext}$ = 114 \si{pA}. In this series, the number of small subthreshold peaks is decreasing with increasing external current, and the distance between the action potentials is decreasing with increasing external current.

For the dynamic simulation of MMOs, it is important to note that the system has multiple time scales. The $s_{1.7}$ variable is by far the slowest variable. Therefore, we perform all the dynamic simulations in this section with a startup phase of 100,000 \si{ms} to achieve the desired asymptotic behavior of all variables. The time window shown, for example, in Fig.~\ref{fig:mmo1}, starts after this startup phase of 100,000 \si{ms}. 

\begin{figure*}
\centering
    \includegraphics[width=\textwidth]{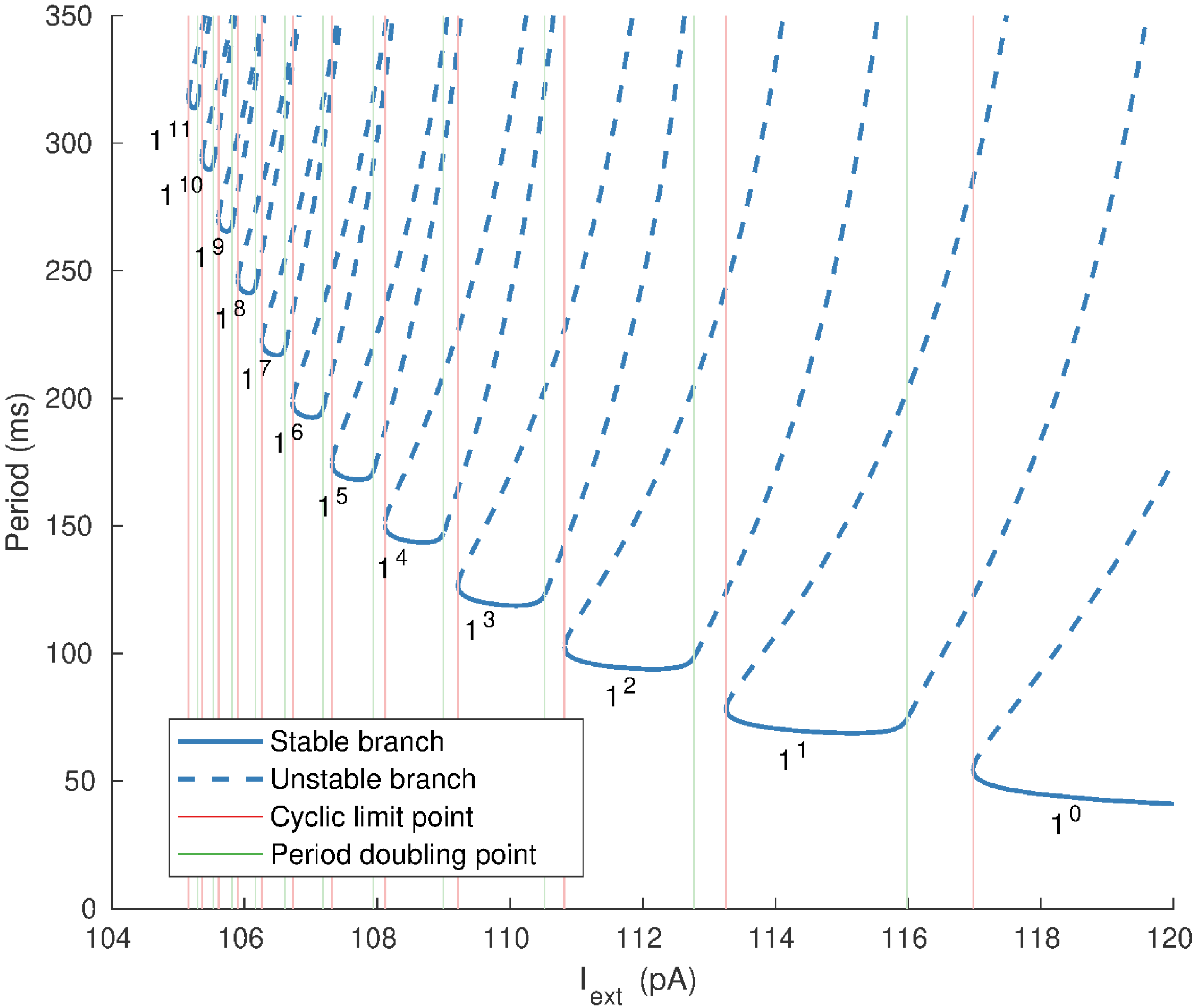}
    \caption{Basic periodic solution branches with one action potential per period in the range of $I_{ext}$ from 105 to 120 \si{pA}. Solid lines: stable periodic solutions, dashed line: unstable periodic solutions.}
    \label{fig:mmobif}
\end{figure*}

To add more details to the picture presented in Fig.~\ref{fig:mmo1}, we perform a one-parameter continuation of the basic MMOs patterns illustrated in Fig.~\ref{fig:mmo1}. Results are shown in Fig.~\ref{fig:mmobif}. The maximum amplitude of these periodic solutions is almost constant and therefore not compelling; instead of the amplitude, we use the period of different solutions for graphical representation of the results.

Fig.~\ref{fig:mmobif} shows a sequence of isolated periodic solution branches, with one action potential per period. The number of small amplitude peaks between the action potentials and the period increases from the right to the left in the direction of decreasing external current. The right most branch with label $1^0$ corresponds to the periodic firing of action potentials without any small amplitude peaks in between, as illustrated in the right diagram of Fig.~\ref{fig:1}a. This periodic solution branch becomes unstable at a cyclic limit point at $I_{ext} = 116.9811$ \si{pA}, which corresponds to the CLP$_3$ point in Fig.~\ref{fig:2}c. On every other branch in Fig.~\ref{fig:mmobif}, the corresponding periodic solution becomes unstable at a cyclic limit point on the left and at a period doubling bifurcation point on the right. The values of $I_{ext}$ at the cyclic limit points are indicated by the red lines and the values at the period doubling bifurcation points by the green lines in Fig.~\ref{fig:mmobif}. These values are also listed in Table~\ref{tab:clppd}.

\begin{table}
\caption{Values of $I_{ext}$ at the cyclic limit points (CLP) and the period doubling bifurcation points (PD) in Fig.~\ref{fig:mmobif}.}
\label{tab:clppd}
\begin{tabular}{lll}
\hline\noalign{\smallskip}
Solution type & CLP (\si{pA})      & PD (\si{pA})       \\
\noalign{\smallskip}\hline\noalign{\smallskip}
$1^{11}$      & 105.1554 & 105.2914 \\
$1^{10}$      & 105.3609 & 105.5277 \\
$1^{9}$       & 105.6055 & 105.8133 \\
$1^8$         & 105.9013 & 106.1649 \\
$1^7$         & 106.2657 & 106.6072 \\
$1^6$         & 106.7246 & 107.1787 \\
$1^5$         & 107.3185 & 107.9409 \\
$1^4$         & 108.1125 & 108.9962 \\
$1^3$         & 109.2164 & 110.5166 \\
$1^2$         & 110.8213 & 112.7732 \\
$1^1$         & 113.2577 & 115.9832 \\
$1^0$         & 116.9811 &          \\
\noalign{\smallskip}\hline
\end{tabular}
\end{table}

\begin{figure*}
\includegraphics[width=1\textwidth]{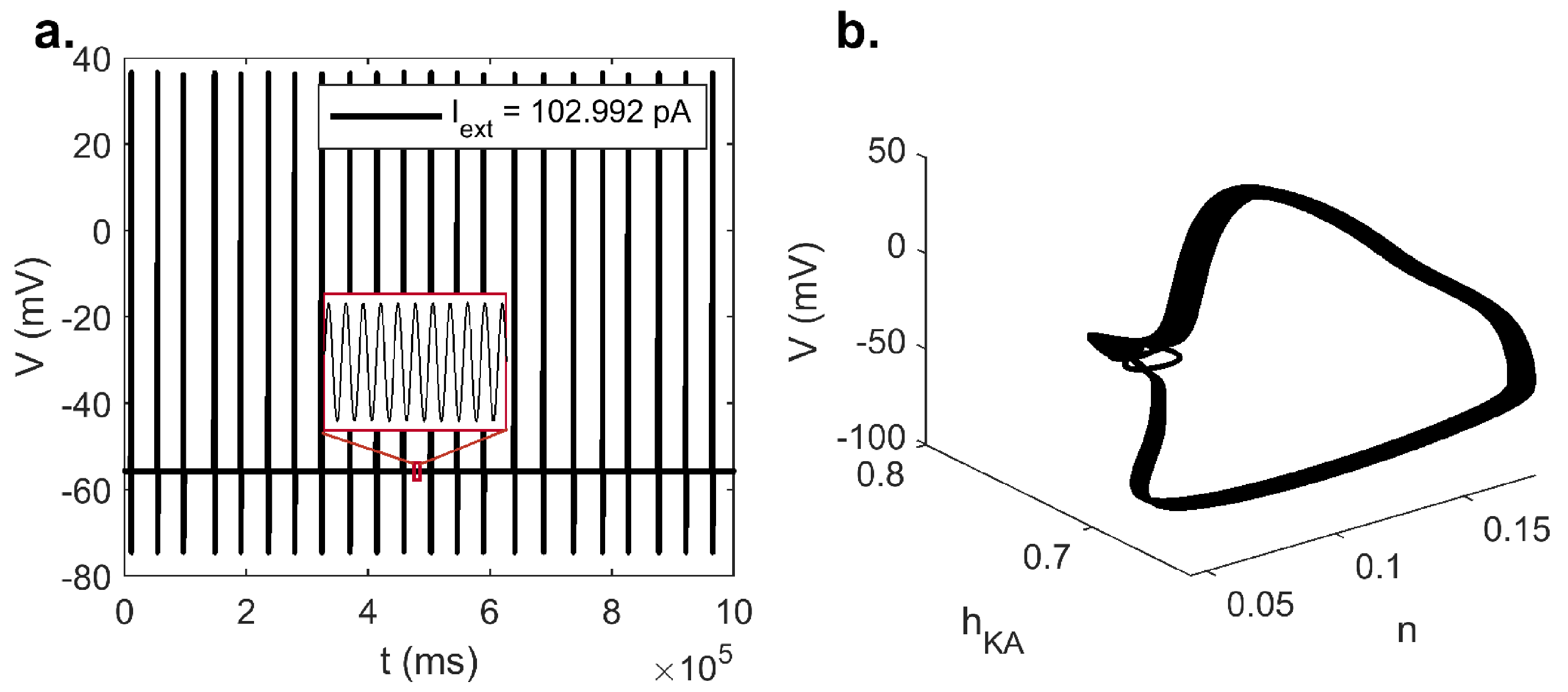}
\caption{a.: MMOs for $I_{ext}$ = 102.992 \si{pA} below the Hopf bifurcation point at $I_{ext}$ = 102.9935 \si{pA}. b.: Representation of the solution in the $V, h_{KA}, n_{K}$ phase diagram.}
\label{fig:Hom}
\end{figure*}

Solutions below 105 \si{pA} are not shown in this figure. Below 105 \si{pA}, the distance between the large amplitude action potentials becomes larger and larger as we increase $I_{ext}$ and finally tends to infinity close to the subcritical Hopf bifurcation point, for which $I_{ext}$ equals 102.9935 \si{pA}. Accordingly, the number of subthreshold peaks becomes larger and larger and their amplitude smaller and smaller as we approach the critical point. As an example, we show dynamic simulations for a value of $I_{ext}=102.992$ \si{pA} which is slightly below the subcritical Hopf bifurcation point, shown in Fig.~\ref{fig:Hom}. The distance between two action potentials is roughly 40,000 \si{ms}. However, as shown in the phase diagram in Fig.~\ref{fig:Hom}b, the orbit is a narrow band and does not seem to be strictly periodic. For $I_{ext}$ slightly below this value, MMOs finally vanish and a stable steady state is obtained.

\begin{figure*}
    \centering
    \includegraphics[width=\textwidth]{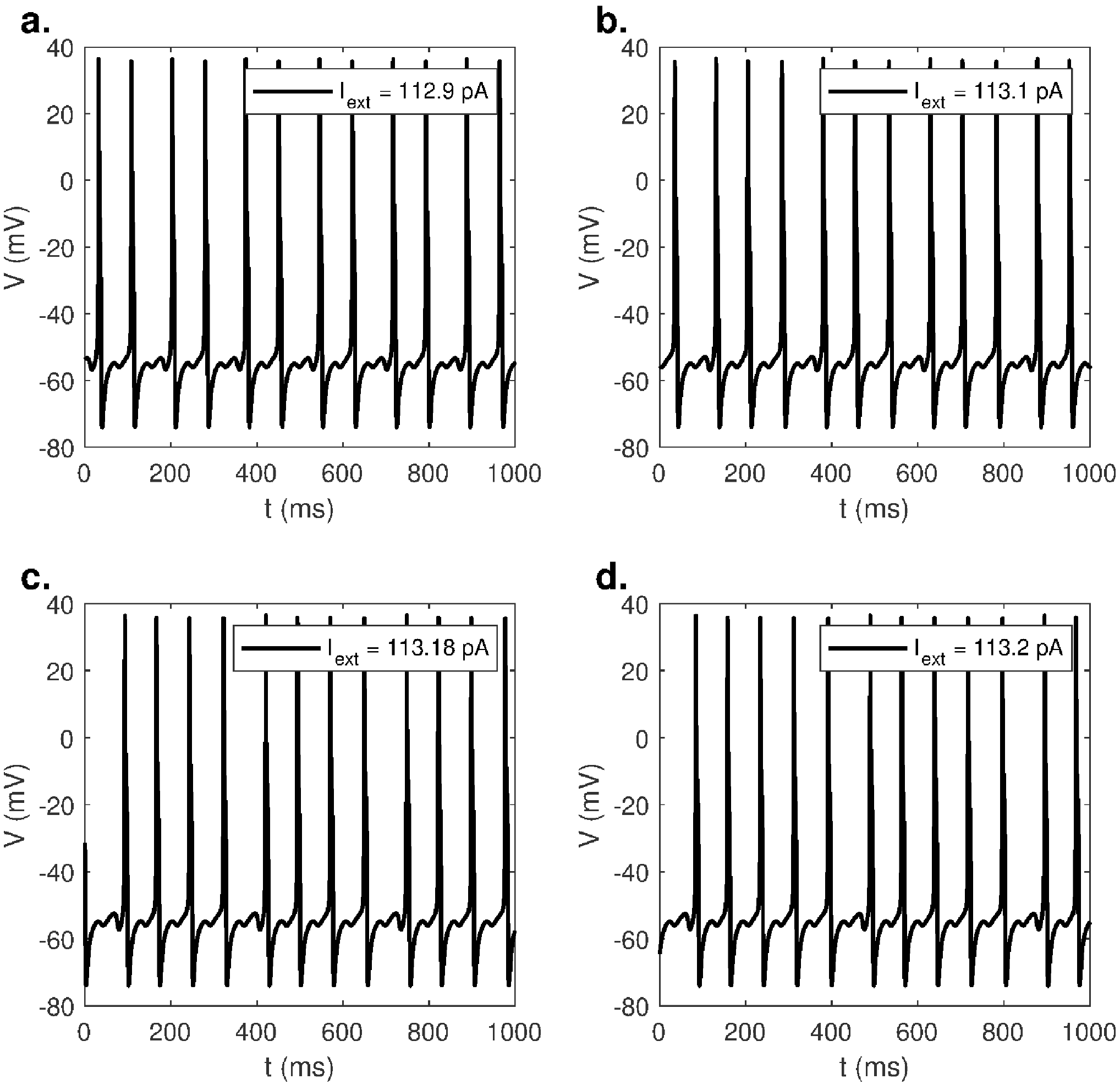}
    \caption{A sequence of concatenated periodic solutions. a.: $1^21^1$ at $I_{ext} = 112.9$ \si{pA}, b.: $1^2(1^1)^2$ at $I_{ext} = 113.1$ \si{pA}, c.: $1^2(1^1)^3$ at $I_{ext} = 113.18$ \si{pA}, d.: $1^2(1^1)^4$ at $I_{ext} = 113.2$ \si{pA}.}
    \label{fig:Fareyex}
\end{figure*}

We find concatenated periodic solutions in the gaps of the basic periodic patterns in Fig.~\ref{fig:mmobif} between the period doubling points and the cyclic limit points of the subsequent solution branches on the right. We further study the dynamic behavior in these regions for selected values of $I_{ext}$ using dynamic simulations. Again, it is crucial to account for the long transient phase introduced by the very slow $s_{1.7}$ variable as described above. Some characteristic patterns of behavior in the range of 112.9 to 113.2 \si{pA} are shown in Fig.~\ref{fig:Fareyex}. According to the aforementioned nomenclature, the solution in Fig.~\ref{fig:Fareyex}a can be characterized as a concatenation between the basic $1^2$ pattern on the left of this value and the basic $1^1$ on the right of this value in Fig.~\ref{fig:mmobif}, leading to a $1^2 1^1$ solution with two action potentials per period. Accordingly, Fig.~\ref{fig:Fareyex}b demonstrates a $1^2(1^1)^2$ pattern with 3 action potentials per period, Fig.~\ref{fig:Fareyex}c a $1^2(1^1)^3$ pattern with 4 action potentials per period, and Fig.~\ref{fig:Fareyex}d a $1^2(1^1)^4$ pattern with 5 action potentials per period.

\begin{figure*}
    \includegraphics[width=\textwidth]{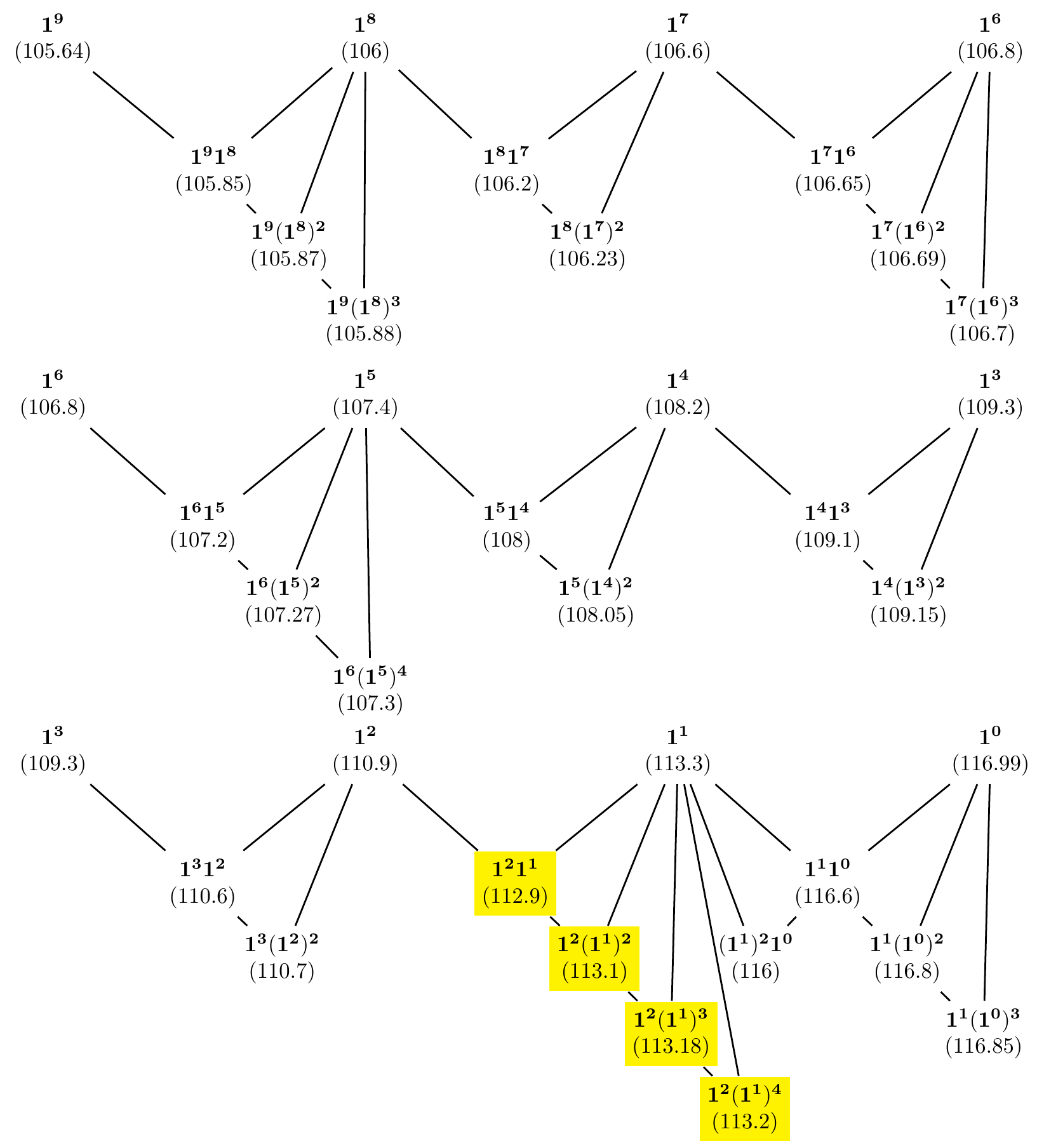}
    \caption{Tree of selected periodic MMOs solutions. Numbers in parentheses are values of $I_{ext}$ in \si{pA} corresponding to the solution on top of it. Solutions highlighted in yellow are shown in Fig.~\ref{fig:Fareyex}.}
    \label{fig:farey}
\end{figure*}

Subsequently, we order the MMOs solutions which were found for selected values of $I_{ext}$ in a tree like structure in Fig.~\ref{fig:farey} containing basic and concatenated MMOs patterns as described above. The corresponding values of $I_{ext}$ in \si{pA} are given in parentheses. The solutions highlighted in yellow correspond to those shown in Fig.~\ref{fig:Fareyex}. It is worth noting that the solution tree is not complete, since only selected values of $I_{ext}$ have been considered. For example, we expect that between the solution $1^6(1^5)^2$ at $I_{ext}$ = 107.27 \si{pA} and the solution $1^6(1^5)^4$ at $I_{ext}$ = 107.3 \si{pA}, another solution of the form $1^6(1^5)^3$ can be found for some suitable value of $I_{ext}$, so that the solutions form a regular so-called Farey sequence~\citep{FareyBZ}. Furthermore, we expect that even higher order concatenated solutions can be found for some suitable values of $I_{ext}$.

\begin{figure*}
    \includegraphics[width=0.8\textwidth]{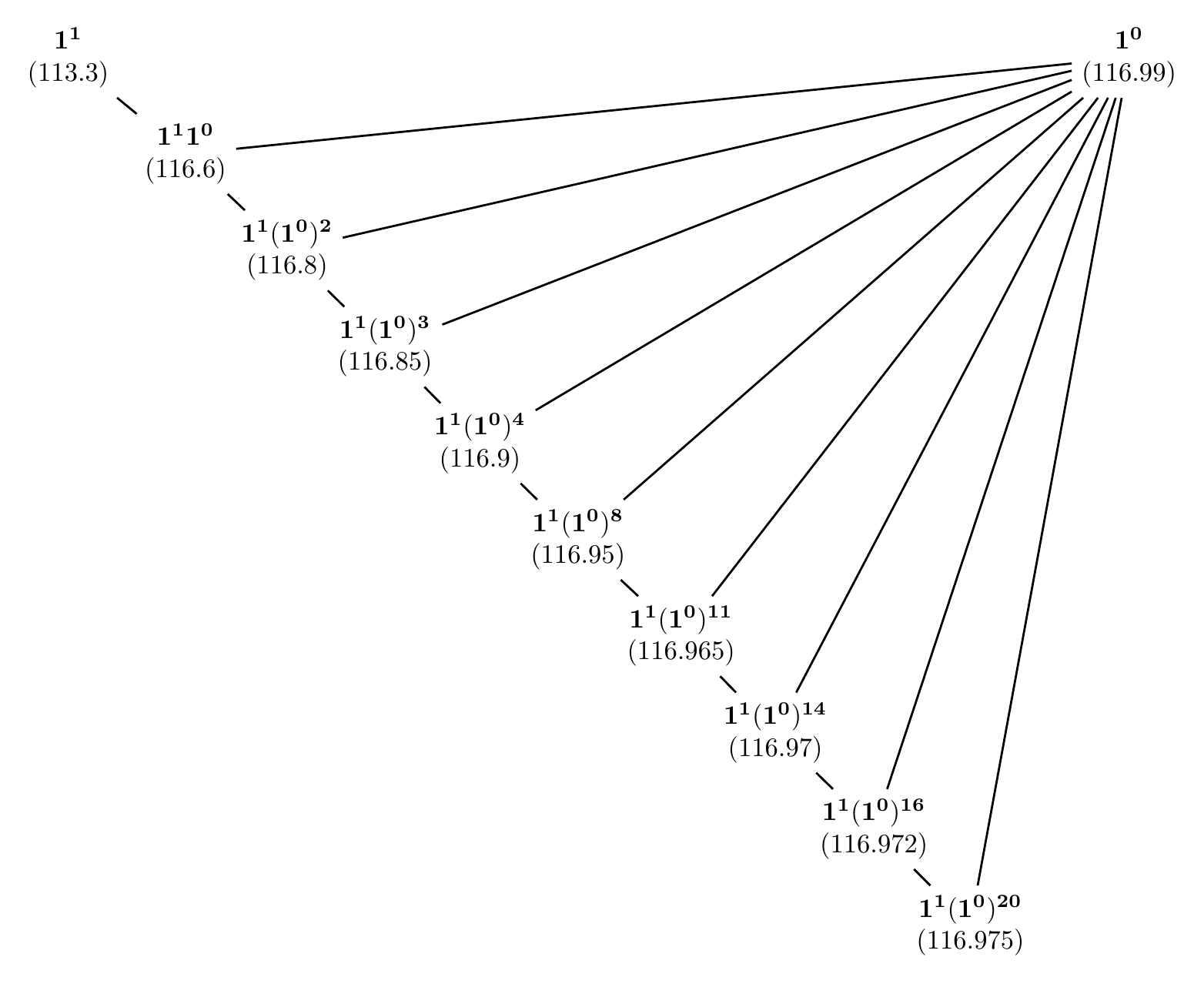}
    \caption{Selected periodic MMOs patterns observed below but close to the cyclic limit point CLP$_3$ in Fig.~\ref{fig:2}c before small amplitude oscillations disappear. Numbers in parentheses are the corresponding values of $I_{ext}$ in \si{pA}, corresponding to the solution on top of it.}
    \label{fig:fareydeep}
\end{figure*}

Close to the cyclic limit point at $I_{ext}$ = 116.9811 \si{pA} in Fig.~\ref{fig:mmobif} corresponding to CLP$_3$ in Fig.~\ref{fig:2}c before the MMOs disappear, the solution consists of one small amplitude peak and multiple large amplitude peaks. If $n$ is the number of large amplitude peaks, this can be written as a concatenation of one $1^1$ solution and $(n-1)$ $1^0$ solutions as $1^1(1^0)^{n-1}$. Selected solutions for this region are shown in Fig.~\ref{fig:fareydeep}. The number of large amplitude action potentials per period is increasing in this sequence from the left to the right.

The solution tree in Fig.~\ref{fig:fareydeep} is also not complete. For example, we expect that one can also find solutions of the form $1^1(1^0)^{12}$ and $1^1(1^0)^{13}$ between $1^1(1^0)^{11}$ and $1^1(1^0)^{14}$ for some suitable value of $I_{ext}$ in between.

For an additional characterization of periodic MMOs, we introduce a firing number $F$. Following \cite{FareyBZ}, $F$ is defined as the number of small amplitude subthreshold peaks per total number of peaks in a period. For a basic $L^S$ pattern, $F$ is given by:
\begin{equation}
    F = \frac{S}{L+S}.
\end{equation}

Accordingly, $1-F$ is the firing rate of action potentials per period, which is even more interesting from the physiological point of view.

The firing number of concatenated MMOs solutions can be calculated using the Farey arithmetic~\citep{hardy1979introduction}. According to this arithmetic, the Farey sum $\oplus$ of two rational numbers $p_1/q_1$ and $p_2/q_2$ is defined as:
\begin{equation}
    \frac{p_1}{q_1} \oplus \frac{p_2}{q_2} = \frac{p_1+p_2}{q_1+q_2}.
\end{equation}

Using this definition, the firing number $F$ of a concatenated solution $L_1^{S_1} L_2^{S_2}$ is, for example, obtained from the following:
\begin{equation}
    F=F_1 \oplus F_2=\frac{S_1+S_2}{L_1+S_1+L_2+S_2}.
\end{equation}

Furthermore, for the firing numbers of two adjacent solutions in a regular Farey sequence $p_1/q_1$ and $p_2/q_2$, the following condition holds:
\begin{equation}
    |p_1 q_2 - p_2 q_1|=1.
\end{equation}

An illustration of the Farey arithmetic for specific MMOs solutions sequence is shown in Table~\ref{tab:FareyAR}.

\begin{table}
\caption{An illustration of MMOs solution sequences satisfying the Farey arithmetic.}
\label{tab:FareyAR}
\begin{tabular}{lll}
\hline\noalign{\smallskip}
MMOs solution & Firing number & $p_1q_2-p_2q_1$  \\
\noalign{\smallskip}\hline\noalign{\smallskip}
$1^{11}$ & $\frac{\displaystyle 11}{\displaystyle 12}$ & \\
\\
$1^{11}1^{10}$ & $\frac{\displaystyle 11}{\displaystyle 12} \oplus \frac{\displaystyle 10}{\displaystyle 11} = \frac{\displaystyle 21}{\displaystyle 23}$ & 1\\
\\
$1^{11}(1^{10})^2$ & $\frac{\displaystyle 21}{\displaystyle 23} \oplus \frac{\displaystyle 10}{\displaystyle 11} = \frac{\displaystyle 31}{\displaystyle 34}$ & 1\\
\\
$1^{11}(1^{10})^3$ & $\frac{\displaystyle 31}{\displaystyle 34} \oplus \frac{\displaystyle 10}{\displaystyle 11} = \frac{\displaystyle 41}{\displaystyle 45}$ & 1\\
\\
$1^{10}$ & $\frac{\displaystyle 10}{\displaystyle 11}$ & 1\\
\\
\noalign{\smallskip}\hline
\end{tabular}
\end{table}

\begin{figure*}
\includegraphics[width=\textwidth]{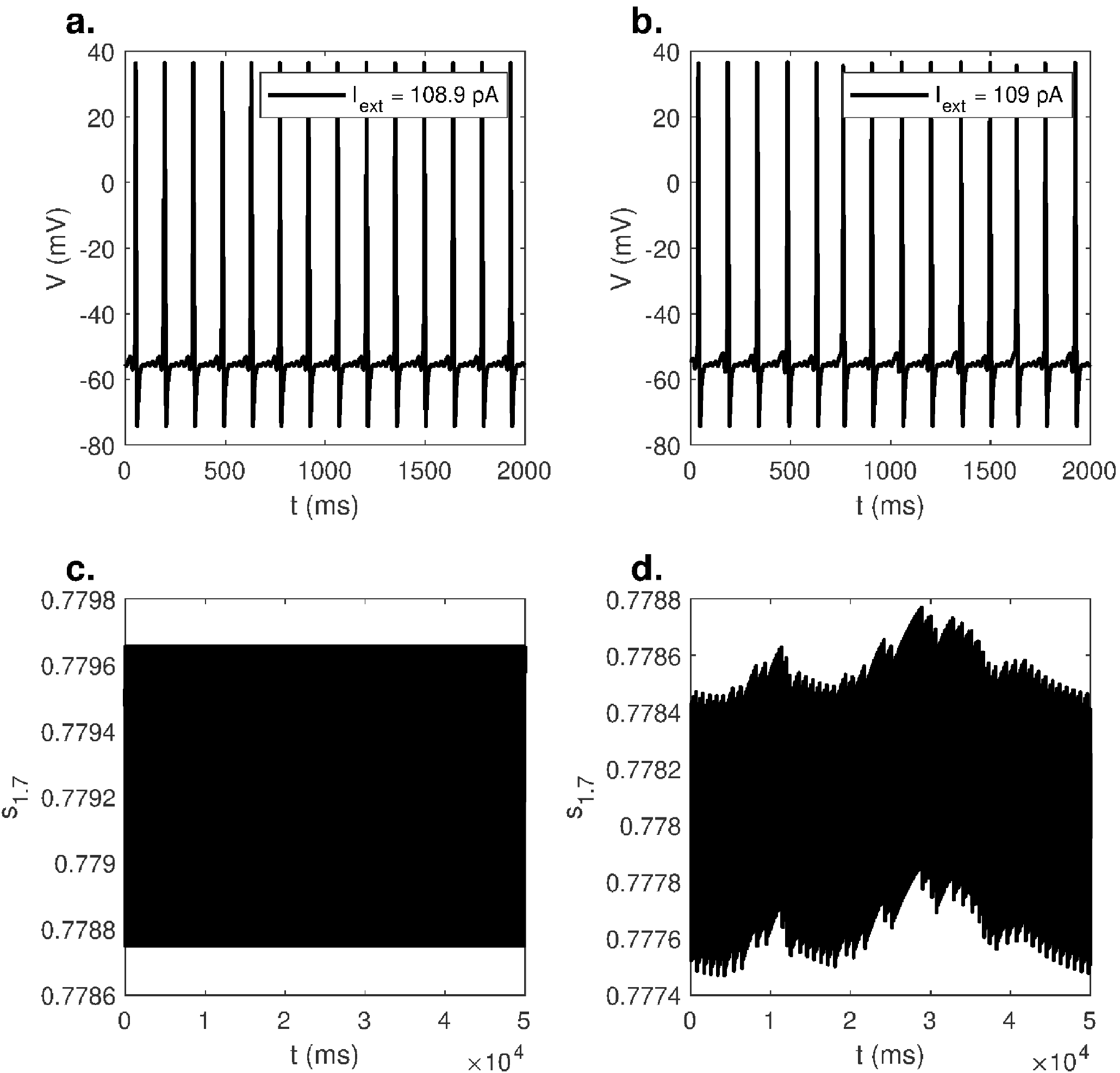}
\caption{Simulations before and after the period doubling bifurcation at $I_{ext}$ = 108.9962 \si{pA}. Left column: $I_{ext} = 108.9$ \si{pA}, right column: $I_{ext} = 109$ \si{pA}. After the period doubling bifurcation, the system exhibits chaotic-like behavior which is evident in the dynamics of $s_{1.7}$.}
\label{fig:chaoscomp}
\end{figure*}

Finally, we also find aperiodic MMOs in the gaps between the stable solution branches in Fig.~\ref{fig:mmobif} for values of $I_{ext}$ slightly above the period doubling points. This is illustrated in Fig.~\ref{fig:chaoscomp} by two simulations. The diagrams on the left demonstrate the dynamic behavior after the startup phase at $I_{ext}$ = 108.9 \si{pA}, below the period doubling point at $I_{ext}$ = 108.9962 \si{pA}, with a stable periodic $1^4$ MMOs solution. The diagrams on the right demonstrate a second solution at $I_{ext}$ = 109 \si{pA} after the startup phase, slightly above the period doubling point. The aperiodic, seemingly chaotic behavior, is not so obvious from the voltage dynamics; however, irregularity is seen in the dynamics of the $s_{1.7}$ variable. In Fig.~\ref{fig:chaoscomp}c, $s_{1.7}$ forms a thick straight band over a long time period, implying that no variation is seen in the $s_{1.7}$ oscillations. However, in Fig.~\ref{fig:chaoscomp}d, irregularity in $s_{1.7}$ is found over this long time period and no repeating patterns are observed.

\section{Discussion}
\label{sec:discussions}
In this work, we attempt to understand the dynamics of a 9D model representative of a small DRG neuron. Small DRG neurons are primary nociceptors and can sense pain. Any damage to them due to injuries, diseases, or genetic disorders, can lead to conditions such as loss or gain of nociceptive pain sensation, and neuropathic or inflammatory pain. A bifurcation analysis of this model can aid in understanding the transition of this system from steady state to mixed-mode oscillations, and finally to full blown periodic firing of action potentials, where oscillations of any frequency indicate pain of a specific form and intensity~\citep{djouhri2006spontaneous,dubin2010nociceptors}.

The model displays rich dynamics, which we investigate by studying the bifurcations numerically using the external applied current as the primary bifurcation parameter and the maximal conductances $\overline{g}_i$ ($i=1.7,1.7,K,KA$) of the sodium and potassium channels as secondary parameters. We show that, in particular, $\overline{g}_{1.8}$ and $\overline{g}_K$ are the most sensitive maximal conductances. We provide a detailed analysis for $\overline{g}_{1.8}$ as the secondary parameter. We show that there is a hard onset of periodic firing of action potentials due to hysteresis between stable steady state and periodic firing of action potentials for low values of $\overline{g}_{1.8}$. This pattern of behavior can also be found in the original Hodgkin-Huxley equations (see, for example, \cite{rinzel/1980a}). For high values of $\overline{g}_{1.8}$, the frequency of firing of action potentials is increasing step by step as we pass through a region of MMOs where the distance between the action potentials is getting smaller and smaller as the number of subthreshold peaks between the action potentials is reduced step by step until they finally vanish. Although the region of MMOs is rather narrow in terms of $I_{ext}$ for the parameter values considered in this paper, it represents a second, and a fundamentally different path to pain.

Using selected dynamic simulations, we conjecture that the periodic MMOs build Farey sequences. Such Farey sequences have also been observed for various other systems displaying MMOs (see~\cite{FareyBZ,albahadily1989farey,hauck1994chaos,andrzej2000farey,Marszalek2012farey} for examples); however, they have not been widely studied for neuron models (see~\cite{krupa2008dopaminergic} for an example). Given the diversity and abundance of neuron models that can generate MMOs~\citep{brons2008mmoreview}, it will be interesting to explore the existence of Farey sequences in other neuron models as well.

Besides periodic, we also find aperiodic MMOs for very small ranges of $I_{ext}$. We conjecture that these aperiodic MMOs solutions are chaotic. Further investigations are required to validate this hypothesis. From a mechanistic point of view, it would be interesting to record such chaotic behavior in DRG cultures and find its implications on pain sensation.

From the mathematical point of view, the 9D model used in this study is rather complex and prohibits further analytical insight as demonstrated for example in~\cite{Rubin2007} for a lower dimensional problem. To gain further theoretical insight it would therefore be desirable to reduce the present model to a lower dimensional problem showing similar patterns of behavior.

From the physiological point of view, the model used in this study is still relatively simple. Towards a more realistic description of small DRG neurons, additional ion channels should be taken into account such as Na\textsubscript{v}1.9, inward rectifier potassium, and calcium channels. Furthermore, this neuron is long and therefore spatially distributed. A more elaborate model needs to be considered in order to perform a further detailed bifurcation analysis and to capture other dynamical behaviours such as bursting~\citep{amir1999drg}. Moreover, in order to build a more realistic model, experimental validation of the observed dynamical patterns along with the current due to each of the ion channels needs to be done, using patch clamp experiments. Lastly, there is vast heterogeneity in the characteristics of action potentials observed in small DRG neuron cultures. While some of these neurons spontaneously fire (repetitive firing at $I_{ext} = 0$ \si{pA}), others do not~\citep{yang2016nav17}. This indicates that simply fixing maximal conductances as constants will not suffice, and there is a need to analyze an ensemble of possible parameter values to capture the heterogeneity observed in electrical
recordings.  

Mathematical understanding of the sensing of pain is necessarily an evolving process of manipulating model scale to suitably match the minimum physiological details associated with pain. Each step in this process involves comparing predictions with experimental observations and identifying how parameters connected with various ion channels relate to pain. Thus both model elaboration and reduction are potential future areas of interest, and can enable a rigorous investigation of possible dynamics that can be displayed by this system. This can further shape our understanding of pain sensation and how it can be controlled.

\appendix
\section{Appendix}\label{sec:appendix}
\subsection{Na\textsubscript{v}1.7 equations}

\begin{equation}
\alpha_{m_{1.7}} = \frac{\displaystyle 15.5}{\displaystyle 1+\mathrm{exp}\left(\frac{\displaystyle (V-5)}{\displaystyle -12.08}\right)}
\end{equation}
\begin{equation}
\beta_{m_{1.7}} = \frac{\displaystyle 35.2}{\displaystyle 1+\mathrm{exp}\left(\frac{\displaystyle V+72.7}{\displaystyle 16.7}\right)}
\end{equation}
\begin{equation}
\alpha_{h_{1.7}}  = \frac{\displaystyle 0.38685}{\displaystyle 1+\mathrm{exp}\left(\frac{\displaystyle V+122.35}{\displaystyle 15.29}\right)}
\end{equation}
\begin{equation}
\beta_{h_{1.7}} = -0.00283 + \frac{\displaystyle 2.00283}{\displaystyle 1 + \mathrm{exp}\left(\frac{\displaystyle (V+5.5266)}{\displaystyle -12.70195}\right)}
\end{equation}
\begin{equation}
\alpha_{s_{1.7}} = 0.00003 + \frac{\displaystyle 0.00092}{\displaystyle 1+\mathrm{exp}\left(\frac{\displaystyle V+93.9}{\displaystyle 16.6}\right)}
\end{equation}
\begin{equation}
\beta_{s_{1.7}} = 132.05 - \frac{\displaystyle 132.05}{\displaystyle 1+\mathrm{exp}\left(\frac{\displaystyle V-384.9}{\displaystyle 28.5}\right)}
\end{equation}
For $x=m_{1.7}, h_{1.7}, s_{1.7}$:
\begin{equation}
    x_{\infty} (V) = \frac{\alpha_x (V)}{\alpha_x (V) + \beta_x (V)},
\end{equation}
and
\begin{equation}
    \tau_x (V) = \frac{1}{\alpha_x (V) + \beta_x (V)} 
\end{equation}

The kinetics of Na\textsubscript{v}1.7 were taken from \cite{sheets2007nav1,choi2011physiological}. $m_{1.7}$ corresponds to the activation gating variable, $h_{1.7}$ to the fast-inactivation gating variable and $s_{1.7}$ to the slow-activation gating variable.     

\subsection{Na\textsubscript{v}1.8 equations}
\begin{equation}
\alpha_{m_{1.8}} = 2.85 - \frac{\displaystyle 2.839}{\displaystyle 1+\mathrm{exp}\left(\frac{\displaystyle V-1.159}{\displaystyle 13.95}\right)}
\end{equation}
\begin{equation}
\beta_{m_{1.8}} = \frac{\displaystyle 7.6205}{1+\mathrm{exp}\left(\frac{\displaystyle V+46.463}{\displaystyle 8.8289}\right)}
\end{equation}
For $x=m_{1.8}$:
\begin{equation}
    x_{\infty} (V) = \frac{\alpha_x (V)}{\alpha_x (V) + \beta_x (V)},
\end{equation}
and
\begin{equation}
    \tau_x (V) = \frac{1}{\alpha_x (V) + \beta_x (V)} 
\end{equation}

\begin{equation}
\tau_{h_{1.8}} = 1.218 + 42.043\times \mathrm{exp}\left(\frac{\displaystyle -(V+38.1)^2}{\displaystyle 2\times15.19^2}\right)
\end{equation}
\begin{equation}
h_{{1.8}_{\infty}} =  \frac{\displaystyle 1}{\displaystyle 1+\mathrm{exp}\left(\frac{\displaystyle V+32.2}{\displaystyle 4}\right)}
\end{equation}

The kinetics of Na\textsubscript{v}1.8 were taken from \cite{sheets2007nav1,choi2011physiological}. $m_{1.8}$ and $h_{1.8}$ are similar activation and inactivation gating variables, respectively.

\subsection{K equations}
\begin{equation}
\alpha_{n_K} = \frac{\displaystyle 0.001265\times (V+14.273)}{\displaystyle 1-\mathrm{exp}\left(\frac{\displaystyle V+14.273}{\displaystyle -10}\right)}
\end{equation}
with $\alpha_{n_K} = 0.001265\times 10$ for $V = -14.273$.
\begin{equation}
\beta_{n_K} = 0.125\times \mathrm{exp}\left(\frac{\displaystyle V+55}{\displaystyle -2.5}\right)
\end{equation}
\begin{equation}
n_{K_{\infty}} = \frac{\displaystyle 1}{\displaystyle 1 + \mathrm{exp}\left(\frac{\displaystyle -(V + 14.62)}{\displaystyle 18.38}\right)}
\end{equation}
\begin{equation}
\tau_{n_K} = \frac{\displaystyle 1}{\displaystyle \alpha_{n_K} + \beta_{n_K}} + 1
\end{equation}

The kinetics of K channel were taken from \cite{schild1994and}. 

\subsection{KA equations}
\begin{equation}
n_{{KA}_{\infty}} = \left(\frac{\displaystyle 1}{\displaystyle 1 + \mathrm{exp}\left(\frac{\displaystyle -(V + 5.4)}{\displaystyle 16.4}\right)}\right)^4
\end{equation}
\begin{equation}
\tau_{n_{KA}} = 0.25 + 10.04\times \mathrm{exp}\left(\frac{\displaystyle -(V + 24.67)^2}{\displaystyle 2 \times 34.8^2}\right)
\end{equation}
\begin{equation}
h_{{KA}_{\infty}} = \frac{\displaystyle 1}{\displaystyle 1 + \mathrm{exp}\left(\frac{\displaystyle V + 49.9}{\displaystyle 4.6}\right)}
\end{equation}
\begin{equation}
\tau_{h_{KA}} = 20 + 50\times \mathrm{exp}\left(\frac{\displaystyle -(V + 40)^2}{\displaystyle 2 \times 40^2}\right)
\end{equation}

The kinetics of KA channel were taken from \cite{sheets2007nav1}

\section*{Acknowledgements}
This project was supported, in part, with support from the Indiana Clinical and Translational Sciences Institute funded, in part by Award Number UL1TR002529 from the National Institutes of Health, National Center for Advancing Translational Sciences, Clinical and Translational Sciences Award. The content is solely the responsibility of the authors and does not necessarily represent the official views of the National Institutes of Health. The authors also thank Dr. Haroon Anwar, New Jersey Institute of Technology, USA, for helping with model selection and building, and for reviewing this manuscript; Max Planck Institute for Dynamics of Complex Technical Systems, Magdeburg, Germany, for sponsoring trips to strengthen the collaboration; and Muriel Eaton and Dr. Yang Yang, Purdue University, USA, for insightful discussions on DRG neurons and pain sensation.

\section*{Conflict of interest}

The authors declare that they have no conflict of interest.


\end{document}